\documentclass[12pt]{article}
\usepackage[a4paper, margin=1in]{geometry}
\usepackage{graphicx} 
\usepackage{amsmath}
\usepackage{booktabs}
\usepackage{multirow}
\usepackage{hyperref}
\usepackage{bm}
\usepackage{array}
\usepackage[symbol]{footmisc}
\usepackage{setspace}
\usepackage{amsfonts}
\usepackage{subcaption}
\usepackage{tikz}
\usetikzlibrary{calc, decorations.pathreplacing, arrows.meta, positioning}
\usepackage{subcaption}

\usepackage[backend=biber,
style=authoryear,
natbib=true,
uniquelist=false     
]{biblatex}

\title{Inference on state occupancy in
covariate-driven hidden Markov models}

\author{}
\date{}

\begin{document}

\maketitle

\begin{center}\vspace{-3em}
    {\large Maya N.\ Vienken$^{1*}$, Jan-Ole Koslik$^1$, Roland Langrock$^1$}
\end{center}

\begin{center}
    $^1$ Statistics and Data Analysis, Department of Business Administration and Economics, Bielefeld University, Bielefeld, Germany
\end{center}

\vspace{1em}

\footnotetext[1]{\texttt{maya.vienken@uni-bielefeld.de}}

\begin{abstract}
    \noindent 
    \begin{enumerate}
    \item Hidden Markov models (HMMs) are natural and popular tools for analysing animal behaviour based on movement, acceleration and other sensor data. In particular, these models allow to infer how the animal's decision-making process interacts with internal and external drivers, by relating the probabilities of switching between distinct behavioural states to covariates.
    \item A key challenge arising in the statistical analysis of behavioural data using covariate-driven HMMs is the models' interpretation, especially when there are more than two states, as then several functional relationships between state-switching probabilities and covariates need to be jointly interpreted. 
    The model-implied probabilities of occupying the different states, as a function of a covariate of interest, constitute a much simpler and hence useful summary statistic. 
    \item A pragmatic approximation of the state occupancy distribution, namely the hypothetical stationary distribution of the model's underlying Markov chain for fixed covariate values, has in fact routinely been reported in HMM-based analyses of ecological data. 
    However, for stochastically varying covariate processes with relatively little persistence, we show that this approximation can be severely biased, hence potentially invalidating ecological inference based on the approximate version of this important summary statistic of interest.
    \item In this contribution, we develop two alternative approaches for obtaining the state occupancy distribution as a function of a covariate of interest --- one based on resampling of the covariate process, the other obtained by regression analysis of the empirical state probabilities. The practical application of these approaches is demonstrated in simulations and a case study on Galápagos tortoise (\textit{Chelonoidis niger}) movement data. 
    \item Our methods enable practitioners to conduct unbiased inference on the relationship between animal behaviour and general types of covariates, thus allowing to uncover the factors influencing behavioural decisions made by animals.  
   \end{enumerate} 
\end{abstract}

\paragraph{Keywords:}{biologging; Markov chain; movement ecology; stationary distribution; statistical ecology; time series analysis}



\section{Introduction}

Hidden Markov models (HMMs) are versatile tools for drawing inference on latent structures from observed time series data. For example, HMMs allow to uncover the dynamics of animal behaviour based on observed movement metrics \citep{mcclintock2020uncovering}, to monitor disease progression based on electronic health records \citep{jackson2003multistate}, or to track learning processes in education \citep{visser2017markov}. 
In such applications, it is often of key interest to learn something about the dynamics of the latent state process, specifically regarding its interaction with internal and external drivers, i.e.\ covariates. In ecology, the overall aim then is to better understand animal decision-making processes as a function of internal conditions and in response to environmental cues \citep{towner2016sex, van2019classifying}. 

To model the effect of covariates on the latent Markovian state process within an HMM, it is common practice to relate the state transition probabilities to linear predictors comprising one or more covariates, using a multinomial logit model for the behavioural decision to be made conditional on the active state \citep{patterson2017statistical, feldmann2023flexible}. 
While conceptually intuitive and appealing with respect to the aim of understanding the decision-making process, such models can quickly become very tedious to interpret. For example, a three-state HMM for animal behaviour --- comprising states such as resting, area-restricted search, and travelling --- involves as many as nine functional relationships between the transition probabilities and \textit{each} covariate considered. 
It is thus desirable to also consult the easier-to-interpret state occupancy distribution implied by the fitted model, i.e.\ the categorical distribution $\Pr(\text{state $i$}|z), i=1,\ldots,N$, of the states being active, as a function of each covariate $z$ of interest.
Since this summary statistic is not readily available, \citet{patterson2009classifying} suggested considering a simple pragmatic
approximation, namely the implied hypothetical stationary distribution of the Markov chain when fixing the covariate at each value of interest. 
This approximation, which has fairly routinely been used in statistical ecology (see, e.g., \citealp{farhadinia2020understanding,byrnes2021evaluating,fahlbusch2022blue,whittington2022towns}) and is also implemented in the R packages \texttt{moveHMM} \citep{michelot2016movehmm} and \texttt{momentuHMM} \citep{mcclintock2018momentuhmm}, corresponds to pretending that the state process follows the same dynamic as implied by a fixed covariate for an extended amount of time, hence ignoring the often highly stochastic variation also in the covariate (e.g.\ temperature, rainfall amount).
Figure 1 illustrates that the bias resulting from ignoring the nature of the covariate process can potentially be very large. 

\begin{figure}[h]
\centering
\includegraphics[width = 0.9\textwidth]{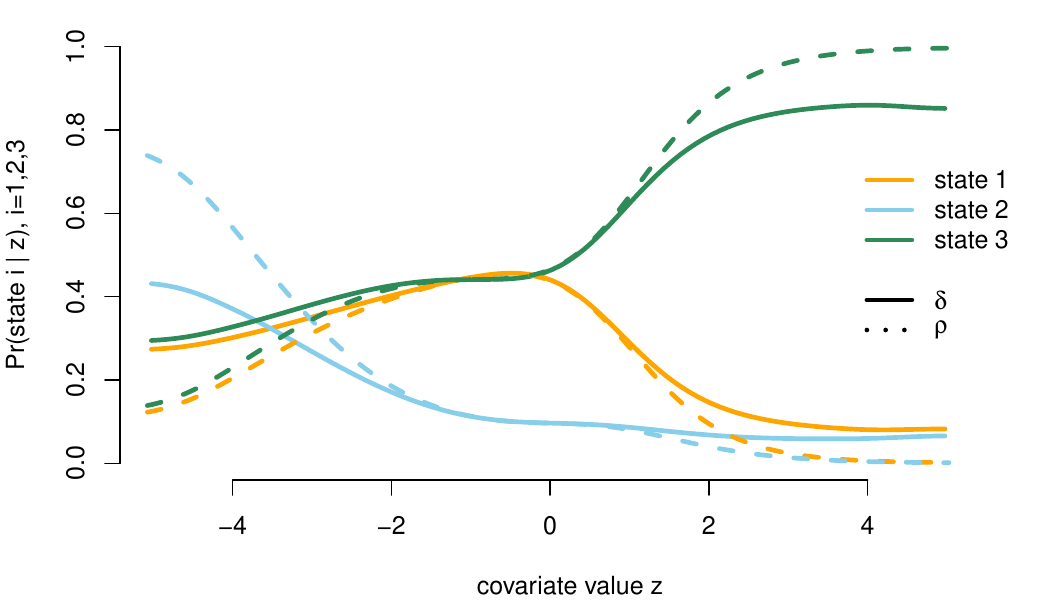}
\caption{True state occupancy distribution ${\bm{\delta}}$ as obtained using Monte Carlo in an example HMM specification (solid lines; see the Online Supplementary Material (S.1) for the parameter values used here), and its approximation by the hypothetical stationary distribution $\bm{\rho}$ as suggested by \citet{patterson2009classifying} (dashed lines). The bias of the latter is largest for the most extreme covariate values, as values in these ranges are typically only short-lived.}
\label{fig:problem}
\end{figure}

For the special case of periodically inhomogeneous Markov chains, most notably including models for time-of-day or seasonal variation, \citet{koslik2023inference} derived an analytical solution for $\Pr(\text{state $i$}|z)$, in that case with $z$ corresponding to the covariate time (of day or year). In that setting, it is possible to arrive at an analytical solution as there is no stochastic variation in the covariate time series. In the general case, with stochastically varying covariates such as temperature, there is no such analytical solution. Instead, $\Pr(\text{state $i$}|z)$ then clearly depends on the nature of the (stochastic) covariate process, and especially its persistence. For example, if the covariate process is very highly persistent over time, then $\Pr(\text{state $i$}|z)$ is the consequence of nearly constant transition dynamics, namely those implied by the transition probability matrix (TPM) when fixing $z$ at the value of interest (such that the approximation by \citealp{patterson2009classifying}, is in fact fairly accurate). If instead the covariate process is more volatile, then the trajectory of the process leading up to the value $z$ of interest becomes crucial, as the associated TPMs together determine the state occupancy probability $\Pr(\text{state $i$}|z)$. In order to arrive at an unbiased version of the summary statistic $\Pr(\text{state $i$}|z)$, we thus need to either complement the HMM analysis by an additional model describing the dynamics of the covariate process or use other methods for estimating the quantity of interest.

In this contribution, we suggest two different types of approaches for arriving at $\Pr(\text{state $i$}|z)$ under a fitted covariate-driven HMM. We will first discuss two methods --- one parametric, the other nonparametric --- that use Monte Carlo resampling of the covariate process to arrive at an estimate of the desired summary statistics. Given the associated risk of not being able to produce realistic covariate trajectories via simulation, we will then explore an alternative approach, where a flexible Dirichlet regression model is fitted to the empirical state probabilities under the fitted model and the associated covariate values. The general performance and practical use will be explored in simulations and a case study involving Galápagos tortoise (\textit{Chelonoidis niger}) movement data.


\section{State occupancy distribution in covariate-driven HMMs}

\subsection{Covariates in the state process of an HMM}
\label{sec:covariate-drivenHMMs}

Hidden Markov models comprise two interdependent stochastic processes: a univariate or multivariate observation process $\{X_t\}_{t \in \mathbb{N} }$ and a latent state process $\{S_t\}_{t\in \mathbb{N}}$. The latter is usually assumed to follow a first-order $N$-state Markov chain.  
The dynamics of $\{S_t\}$ are then governed by the initial state distribution $\bm{\delta}^{(1)}=\bigl(\text{Pr}(S_1=1), \ldots, \text{Pr}(S_1=N)\bigr)$ and the state transition probabilities $\gamma_{ij}^{(t)}=\Pr(S_t=j|S_{t-1}=i)$, which are summarised in the $N \times N$ TPM $\bm{\Gamma}^{(t)} = (\gamma_{ij}^{(t)})$.
At each time $t$, the system occupies a latent state $S_t \in \{1,..., N\}$ that determines which of $N$ component distributions generates the observation $X_t$. The observations are assumed to be conditionally independent of each other, given the states. For more comprehensive introductions to HMMs, we refer to \citet{zucchini}, \citet{mcclintock2020uncovering}, and \citet{mews2025build}. 

\begin{figure}[h]
\centering
\begin{tikzpicture}
    \coordinate (A) at (0,0);
    \coordinate (B) at (2,0);
    \coordinate (C) at (4,0);
    \coordinate (D) at (6,0);
    \coordinate (E) at (8,0);

    \filldraw[fill=black!5, thick] (A) circle (0.5);
    \draw (A) node {$\dots$};
    \filldraw[fill=black!5, thick] (B) circle (0.5);
    \draw (B) node {$S_{t-1}$};
    \filldraw[fill=black!5, thick] (C) circle (0.5);
    \draw (C) node {$S_{t}$};
    \filldraw[fill=black!5, thick] (D) circle (0.5);
    \draw (D) node {$S_{t+1}$};
    \filldraw[fill=black!5, thick] (E) circle (0.5);
    \draw (E) node {$\dots$};
    \draw[-{Latex[length=2mm]}] ($(A)+(0.52,0)$) -- ($(B)-(0.52,0)$);
    \draw[-{Latex[length=2mm]}] ($(B)+(0.52,0)$) -- ($(C)-(0.52,0)$);
    \draw[-{Latex[length=2mm]}] ($(C)+(0.52,0)$) -- ($(D)-(0.52,0)$);
    \draw[-{Latex[length=2mm]}] ($(D)+(0.52,0)$) -- ($(E)-(0.52,0)$);
    \filldraw[fill=black!5, thick] ($(B)+(0,1.5)$) circle (0.5);
    \draw ($(B)+(0,1.5)$) node {\textcolor{black}{$X_{t-1}$}};
    \filldraw[fill=black!5, thick] ($(C)+(0,1.5)$) circle (0.5);
    \draw ($(C)+(0,1.5)$) node {\textcolor{black}{$X_{t}$}};
    \filldraw[fill=black!5, thick] ($(D)+(0,1.5)$) circle (0.5);
    \draw ($(D)+(0,1.5)$) node {\textcolor{black}{$X_{t+1}$}};
    \draw[-{Latex[length=2mm]}, color=black] ($(B)+(0,0.52)$) -- ($(B)+(0,1.5)-(0,0.52)$);
    \draw[-{Latex[length=2mm]}, color=black] ($(C)+(0,0.52)$) -- ($(C)+(0,1.5)-(0,0.52)$);
    \draw[-{Latex[length=2mm]}, color=black] ($(D)+(0,0.52)$) -- ($(D)+(0,1.5)-(0,0.52)$);

    \filldraw[fill=black!5, thick] ($(A)-(-1,1.8)$) circle (0.5);
    \draw ($(A)-(-1,1.8)$) node {${z}_{t-1}$};
    \filldraw[fill=black!5, thick] ($(B)-(-1,1.8)$) circle (0.5);
    \draw ($(B)-(-1,1.8)$) node {${z}_{t}$};
    \filldraw[fill=black!5, thick] ($(C)-(-1,1.8)$) circle (0.5);
    \draw ($(C)-(-1,1.8)$) node {$ {z}_{t+1}$};
    \filldraw[fill=black!5, thick] ($(D)-(-1,1.8)$) circle (0.5);
    \draw ($(D)-(-1,1.8)$) node {${z}_{t+2}$};
    \draw ($(A)+(1,-.6)$) node {$\bm{\Gamma}^{(t-1)}$};
    \draw ($(B)+(1,-.6)$) node {$\bm{\Gamma}^{(t)}$};
    \draw ($(C)+(1,-.6)$) node {$\bm{\Gamma}^{(t+1)}$};
    \draw ($(D)+(1,-.6)$) node {$\bm{\Gamma}^{(t+2)}$};
 \draw[-{Latex[length=2mm]}] ($(A)+(1,-1.8)+(0,0.52)$) -- ($(A)+(1,-0.8)$);
 \draw[-{Latex[length=2mm]}] ($(B)+(1,-1.8)+(0,0.52)$) -- ($(B)+(1,-0.8)$);
 \draw[-{Latex[length=2mm]}] ($(C)+(1,-1.8)+(0,0.52)$) -- ($(C)+(1,-0.8)$);
 \draw[-{Latex[length=2mm]}] ($(D)+(1,-1.8)+(0,0.52)$) -- ($(D)+(1,-0.8)$);
\end{tikzpicture}
\caption{Dependence structure of a covariate-driven HMM. The covariate value $z_t$ at time $t$ influences the TPM $\bm{\Gamma}^{(t)}$ governing the transition from time $t-1$ to time $t$.}
\label{fig:covariatestpm}
\end{figure}
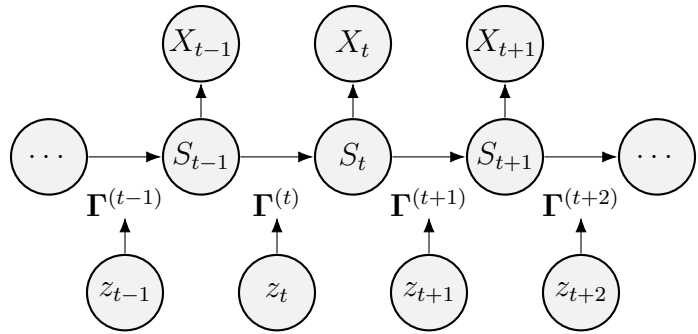

In our setting, illustrated in Figure \ref{fig:covariatestpm}, covariates affect the state transition probabilities and thereby shape the dynamics of the system, i.e.\ the way an animal makes behavioural decisions depending on internal conditions and external effects. The probability of transitioning from state $i$ to state $j$, $i,j=1,\ldots,N$, at time $t$ is then modelled using row-wise application of a (multinomial) logit link to the entries of the TPM. This ensures $\gamma_{ij}^{(t)} \in [0,1]$ and row sums of one. Specifically, 
\begin{equation}
 \gamma_{ij}^{(t)} = \gamma_{ij}(\mathbf{z}_t)= \frac{e^{\eta_{ij}^{(t)}}}{\sum_{k=1}^{N} e^{\eta_{ik}^{(t)}}}, \quad \text{with} \quad \eta_{ij}^{(t)} = \begin{cases}
\beta_0^{(ij)} + \sum^{P}_{p=1} \beta_p^{(ij)} z_{t,p} & \text{if } i\ne j; \\
0 & \text{otherwise}.
\end{cases}
\label{eq:multilogitlinkTPM}
\end{equation}
While we usually use index notation for $\bm{\Gamma}^{(t)}$ to stress its inhomogeneity, we occasionally use functional notation, i.e.\ $\bm{\Gamma}(z_t)$, to stress that this inhomogeneity is induced by the TPM being a function of time-varying covariates.

In terms of formulating the model, the inclusion of covariates via row-wise multinomial logistic regression within the TPM is fairly straightforward. However, the interpretation of such a model can be very tedious, as Equation (\ref{eq:multilogitlinkTPM}) involves $N\cdot (N-1)$ non-zero linear predictors ($N-1$ for each row in the TPM). In many HMM analyses of animal movement data, models with $N=3$ (or more) states are fitted (\citealp{grecian2018understanding,creel2020hidden,beumer2020application}), such that corresponding model formulations with a covariate in the state process involve nine functional relationships between state transition probabilities and the covariate of interest (of which strictly speaking only six are freely estimated, due to the row sum constraints); see \ref{fig:CaseStudyTranstionprob} for an illustration. 
It is thus desirable to also be able to present a simpler summary statistic, namely the probabilities of occupying the different states as a function of the covariate, i.e.\ $\Pr(\text{state $i$}|z)$. This substantially simplifies the interpretation, as one then needs to make sense of only $N$ such state occupancy probabilities, as opposed to $N^2$ probabilities of switching between a pair of states.

\subsection{Hypothetical stationary distribution}
\label{sec:hypotheticalstadist}
\citet{patterson2009classifying} observed the tediousness of interpreting multiple linear predictors linked to the TPM, and suggested a pragmatic solution to approximate the easier-to-interpret implied state occupancy probability $\Pr(\text{state $i$}|z)$. Specifically, they suggested to calculate the hypothetical stationary (or equilibrium) distribution of the state process when fixing the covariate at some level $z_0$. For the associated (constant) TPM $\bm{\Gamma}(z_0)$, the stationary distribution $\bm\rho (z_0)$ is then calculated as the solution to $\bm\rho (z_0) = \bm\rho (z_0)\bm{\Gamma}(z_0)$ subject to $\sum_{j=1}^{N} \rho_j(z_0)=1$.
By repeating this computation over a fine grid of plausible covariate values, a function $\rho_i (z)$ is obtained, which serves as an approximation to $\Pr(\text{state $i$}|z)$, for each state $i=1,\ldots,N$. In case of multiple covariates being incorporated into the transition probabilities, the function $\rho_i (z)$ for the covariate $z$ of interest is typically obtained fixing the remaining covariates at their respective means.

The hypothetical stationary distribution $\bm\rho (z)$ will often be a useful approximation and has indeed routinely been used in practice (see, e.g., \citealp{farhadinia2020understanding,byrnes2021evaluating,klappstein2023flexible}). However, the accuracy of this approximation is highly dependent in particular on the persistence of the covariate process. In settings where the covariate is highly persistent, the hypothetical stationary distribution will be fairly accurate, as the covariate values change only slowly over time, such that the hypothetical stationary model used to calculate  $\bm\rho (z)$ is in fact similar to the true data-generating process. 
However, in settings with less persistence in the covariate process, i.e.\ more rapid changes in the values of the covariate over time, there is a larger discrepancy between the hypothetical stationary model and the true process. As a consequence, the approximation by the hypothetical stationary distribution will often be poor, as the probabilities of occupying a state at any given time $t$ --- and hence the implied overall state occupancy probability --- highly depend on the preceding state dynamics determined by $\bm{\Gamma}^{(t-1)}$, $\bm{\Gamma}^{(t-2)}$, $\bm{\Gamma}^{(t-3)}, \ldots$, which in turn are functions of the (volatile) time-varying covariate and hence will be anything but constant (cf.\ Figure \ref{fig:problem}).
In fact, for given parameters of the predictors determining the TPM, the hypothetical stationary distribution will be the same regardless of the persistence of the covariate processes, when in fact the actual state occupancy probability strongly depends on the nature of the covariate processes. 

Throughout this paper, we use $\Pr(\text{state $i$}|z)$ for notational simplicity to denote the quantity of interest, i.e.\ the state occupancy probability as a function of a covariate $z$. Strictly speaking, for large $t$, we consider
\begin{equation}
\label{eqn:int_statedist}
\Pr(S_t = i \mid Z_t = z) = \mathbb{E} \bigl[ \Pr(S_t = i \mid Z_1, \dots, Z_t) \mid Z_t = z \bigr],
\end{equation}
where the expected value is taken with respect to the unknown conditional distribution of $Z_1, \ldots, Z_{t-1} \mid Z_t = z$.\footnote{Equation \eqref{eqn:int_statedist} can be obtained by applying the law of iterated expectations to $\bm{1}_{S_t = i}$.}
We assume stationarity of the covariate process, such that the specific $t$ is not relevant.
Equation (\ref{eqn:int_statedist}) can then be viewed as a mapping $z \mapsto \Pr(S_t = i \mid Z_t = z)$, which is our summary statistic of interest.
For a fixed covariate sequence, evaluating the quantity inside the expectation is straightforward: it simply reduces to forming the vector-matrix product of the initial state distribution with the covariate-dependent TPMs $\bm{\Gamma}(z_1),\ldots,\bm{\Gamma}(z_{t-1})$.
However, to obtain an unbiased estimate of \eqref{eqn:int_statedist}, information on the temporal dynamics of the covariate process preceding the value $z$ --- i.e.\ its conditional distribution --- needs to be taken into account.
In practice, neither do we know the true distribution and dynamics of the covariate process, nor would we, in general, be able to compute the expected value given this knowledge. 
Consequently, we settle for methods to rigorously approximate \eqref{eqn:int_statedist} based on Monte Carlo approaches (Section \ref{sec:resampling-approaches}) or the sample of the covariate process already available (Section \ref{sec:directmodelling}).

\subsection{Resampling-based methods}
\label{sec:resampling-approaches}

One natural approach to include information on the covariate process is to use resampling methods. Specifically, using resampling techniques --- the details of which are provided below --- we can generate very long simulated trajectories of the covariate process with temporal dynamics similar to those of the empirical covariate process. We can then calculate the state occupancy probabilities $\bm\delta^{(t)}$ as implied by the fitted HMM, for each $t$ in the simulated covariate trajectory, as
\begin{equation}
\bm\delta^{(t)} 
= \bm\delta^{(1)} \cdot \bm\Gamma(z_2) \cdot \bm\Gamma(z_3) \cdot \ldots \cdot \bm\Gamma(z_t).
\label{eq:statedistcov}
\end{equation} 
Our summary statistic of interest, the overall state occupancy probability $\Pr(\text{state $i$}|z)$ for a given covariate value $z$, can then be approximated as follows. First, we divide the range of covariate values into a moderately large number (e.g.\ 50) of equidistant intervals. Second, for each of those intervals, we identify all time points $t$ from the simulated covariate trajectory for which the associated covariate value falls into that interval. Third, again for each interval, we calculate the mean of all corresponding $\delta_i^{(t)}$ as our approximation for $\Pr(\text{state $i$}|z)$ for all $z$ from that interval. This binning approach corresponds to computing the state distribution based on conditional samples of the covariate history, given it ended up in a specific bin.

If the covariate sequence is simulated from the actual underlying data-generating process, then the approximation is due to Monte Carlo error only and can hence be made arbitrarily accurate by increasing the length of the simulated covariate sequence.
It will usually be necessary to simulate long covariate sequences, typically much longer than the observed sequence, in order to make sure that the effective range of covariate values is sufficiently well covered.  
In particular, the most extreme covariate values are encountered only rarely, limiting our ability to infer $\Pr(\text{state $i$}|z)$ reliably.

The actual generation of the simulated covariate sequences can be accomplished in several different ways, two of which --- autoregressive modelling and block bootstrapping --- we have identified as particularly promising and hence present in the following\footnote{We also tested generative adversarial networks \citep{goodfellow2014generative, dahl2022time}, but ultimately decided not to pursue this approach further, as we favour a transparent and easy-to-implement (in R) approach.}. Autoregressive (AR) processes are appropriate when the covariate process linearly depends on its previous values \citep{ives2010analysis}, 
$$Z_t = \sum^{p}_{i=1}\phi_iZ_{t-i}+\epsilon_t,$$ 
with the coefficients $\phi_1,\ldots,\phi_p$ and the variance $\sigma^2$ of the normally distributed error term $\epsilon_t$ to be estimated. The order $p$ is typically chosen using model selection criteria, e.g.\ the AIC. In case the transition probabilities depend on multiple covariates, vector-autoregressive (VAR) processes can be used to fully capture the dynamics of the multivariate covariate time series. Using (V)AR processes to generate artificial covariate sequences will typically work well whenever the dynamics of the covariate process are relatively simple. However, in more complex situations, e.g.\ with periodic variation (diel or seasonal) or a highly skewed covariate distribution, it will usually be preferable to use a more flexible, model-agnostic approach to avoid the possible model misspecification and resulting error in the approximation of $\Pr(\text{state $i$}|z)$.

Block bootstrapping (BB) \citep{kunsch1989jackknife} is a resampling technique designed to account for serial dependence, and specifically the local correlation patterns that are pertinent for the calculation of $\Pr(\text{state $i$}|z)$. 
To implement the block bootstrap, a block length $L$ must be chosen that adequately captures the dependence structure in the data, avoiding values that are too small as to retain the local dependence structure but also values that are larger than necessary as otherwise the simulated series becomes too similar to the observed one. If the covariate series exhibits a periodic pattern, $L$ should match one full seasonal cycle, such as $L = 24$ for hourly observations with a daily pattern (cf.\ Section \ref{sec:CaseStudy}).
The covariate series of length $T$ is then divided into $M = T/L$ consecutive, non-overlapping blocks. The blocks are defined as $B_i = (Z_{(i-1)L+1}, Z_{(i-1)L+2}, \ldots, Z_{iL})$ for $i = 1, \ldots, M$.
From this collection of blocks, we draw $M^*$ blocks with replacement, with $M^* \gg M$. The sampled blocks $B^*_{i1}, B^*_{i2}, \ldots, B^*_{iM^*}$ are concatenated to form a bootstrapped time series. This approach naturally extends to multivariate covariate time series.

Both resampling approaches are suited to scenarios with relatively short time series, where there is limited information on the empirical covariate distribution and the dynamics of the covariate process, hindering the approximate calculation of $\Pr(\text{state $i$}|z)$. If, however, the observed time series is in fact relatively long, then there will typically be sufficient such information, in which case the simulation of artificial covariate series will not be necessary. Furthermore, there will sometimes be practical settings in which the generation of artificial covariate sequences will be very difficult, in particular if the covariate of interest depends on the observed time series, as will often be the case in statistical ecology (e.g.\ distance to water depends on the spatial location of an animal). The approach presented in the subsequent section is suitable for both of these scenarios and constitutes a user-friendly and likely less error-prone alternative to the resampling-based approaches.

\subsection{Flexible Dirichlet regression}
\label{sec:directmodelling}

Suppose that we are faced with an at least moderately long observed time series and associated covariate sequence. We can then calculate and visualise as a scatterplot the model-implied state occupancy probabilities $\bm\delta^{(t)}$ from Equation (\ref{eq:statedistcov}) for the observed covariate values (cf.\ Figures \ref{fig:HypotheticalCS}--\ref{fig:flexibledirichletRegS}). These quantities are effectively empirical versions of the desired function $\Pr(\text{state $i$}|z)$, subject to the stochasticity in the covariate process. Conceptually, it thus makes sense to simply smooth these scatterplot data, i.e.\ to fit a regression curve to $\bm\delta^{(t)}$ regarded as function of $z_t$ that we then take as an approximation of the summary statistic of interest, $\Pr(\text{state $i$}|z)$, for $i=1,\ldots,N$.


Since $\bm{\delta}^{(t)} = (\delta_{1}^{(t)}, \ldots, \delta_{N}^{(t)})$ represents compositional data, i.e.\ 
$\delta_i^{(t)} \in (0, 1)$ for $i=1, \ldots, N$ subject to $\sum_{i=1}^N \delta_i^{(t)} = 1$,
it is natural to assume that $\bm{\delta}^{(t)} \sim \text{Dirichlet}(\bm{\alpha}_t)$.
We then model the vector of positive concentration parameters, $\bm\alpha_t = (\alpha_{t1}, \ldots, \alpha_{tN})$, as a smooth function of the covariate $z_t$. Although the Dirichlet distribution belongs to the exponential family, its normalising constant $\mathrm{B}(\bm\alpha)$ incorporates all components, which prevents the use of the standard generalised additive model (GAM) framework, designed for independent scalar responses. We therefore adopt the GAM principle of representing smooth effects with penalised splines but fit a Dirichlet regression model that explicitly handles the joint likelihood of the compositional response \citep{douma2019analysing}.
The smooth effects are defined as $\alpha_{ti}(z_t)=\text{exp}\{f_i(z_t)\}$, $i=1,\ldots,N$,
where $f_i(z_t)$ is a spline-based smooth function penalised through a quadratic smoothing term on its coefficients. Model fitting proceeds by maximising a penalised log-likelihood, where the penalty controls the smoothness of $f_i(\cdot)$. Both the spline coefficients and the smoothing parameters are estimated jointly using quasi-restricted maximum likelihood, treating the smooth terms as random effects \citep{koslik2024efficientsmoothnessselectionnonparametric}. This results in a smooth Dirichlet regression model, an extension of classical Dirichlet regression that replaces linear predictors with smooth, data-driven functions. The fitted model provides estimates of $\Pr(\text{state $i$}|z)$, while preserving the compositional structure of the state probabilities.

\section{Simulation experiments}
\label{sec:simulation-experiments}

We carry out simulation experiments to investigate the performance of the resampling-based approaches described in Section \ref{sec:resampling-approaches}. In the interest of brevity, information on the performance of the approach based on Dirichlet regression, as described in Section \ref{sec:directmodelling}, is illustrated in the Online Supplementary Material (S.2). 
For the resampling-based approaches, we designed a series of simulation experiments using three different model specifications: (I) a 3-state Gaussian HMM with covariate-driven transition probabilities, where the covariate follows a highly persistent AR(1)  process ($\phi = 0.95$); (II) a version with a moderately persistent AR(1) covariate process ($\phi = 0.7$); and (III) a version where the covariates are in fact not generated from an AR(1) process, but rather exhibit a periodic pattern generated by a trigonometric function plus noise (for details, see the R code provided in the Online Supplementary Material (S.1)).

\begin{figure}[!htb]
    \centering
    \includegraphics[width=1\linewidth]{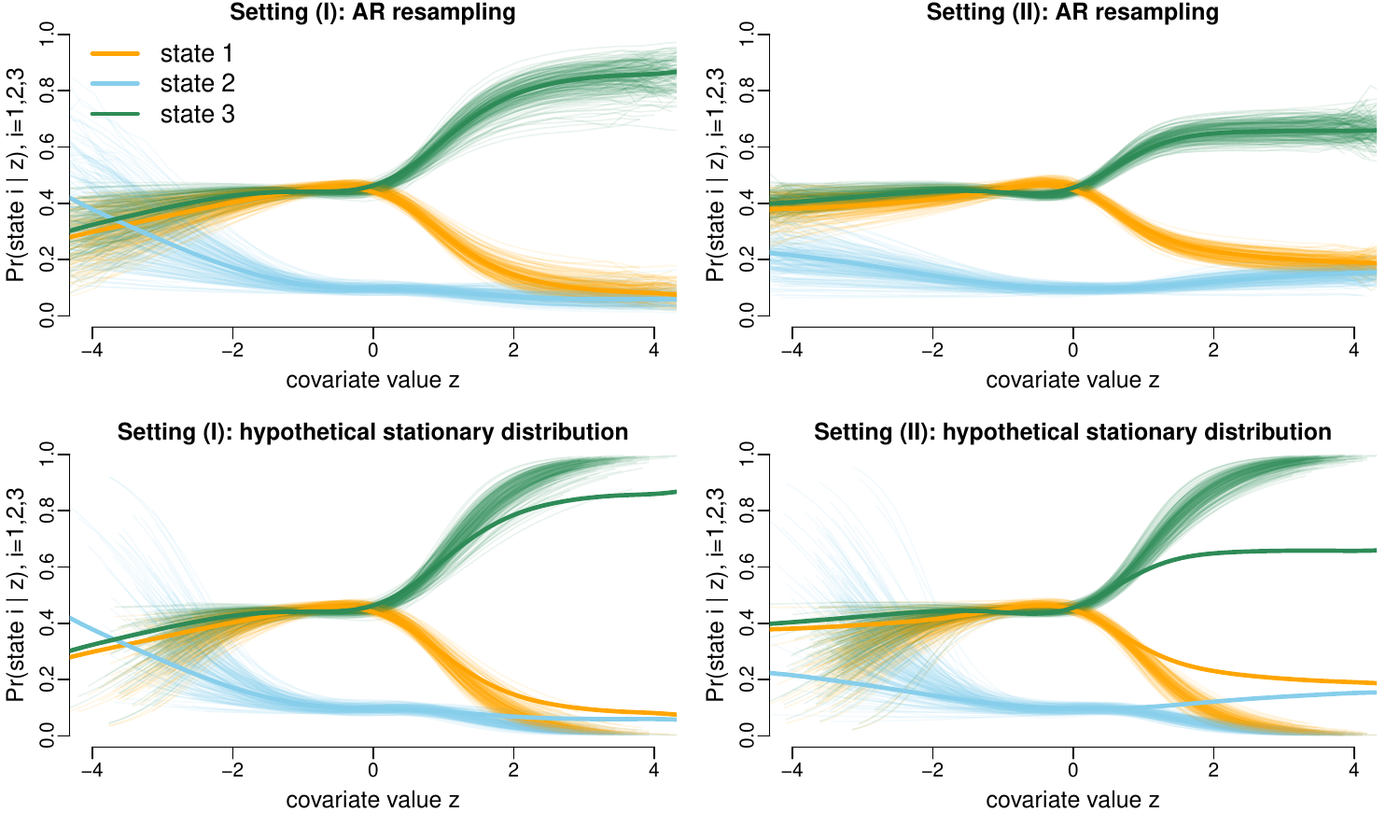}
    \caption{The thin lines represent the estimated covariate-dependent state occupancy distributions based on HMMs fitted to 200 simulated datasets, under Settings (I) and (II) (left panels: highly persistent covariate; right panels: moderately persistent covariate). The upper plots show the results under the AR resampling approach, while the lower plots display the hypothetical stationary distributions. In all panels, the thicker lines represent the true state occupancy distribution obtained using Monte Carlo.}
    \label{fig:simstudyARHyp}
\end{figure}

For each scenario, we first determine the true functions $\Pr(\text{state $i$}|z)$, $i=1,\ldots,N$ --- which are not analytically available --- by Monte Carlo, i.e.\ by simulating an extremely long covariate time series. Using the binning procedure already described in Section \ref{sec:resampling-approaches}, we obtain $N=3$ continuous curves by connecting the function values from neighbouring bins. These curves represent the target quantities we aim to estimate, subject to a negligible Monte Carlo error.

Next, we set up a code routine that repeats the following process 200 times, for each of the three model setups. For each repetition, we simulate data of length $T=2000$, 
then using numerical maximum likelihood \citep{mcclintock2020uncovering} to fit the correctly specified 3-state covariate-driven Gaussian HMM to the simulated data. Based on the resulting parameter estimates, we apply the resampling-based approaches to approximate the summary statistic of interest, $\Pr(\text{state $i$}|z)$, $i=1,\ldots,N$. 

Figure \ref{fig:simstudyARHyp} displays the 200 estimates of $\Pr(\text{state $i$}|z)$ obtained under the HMMs fitted to the 200 datasets simulated under Settings (I) and (II). The lower plots show the approximation of $\Pr(\text{state $i$}|z)$ based on the hypothetical stationary distribution, while the upper plots show the results based on the AR resampling approach under the same fitted models. For Setting (I), with the highly persistent covariate, the hypothetical stationary distribution shows discrepancies to the true $\Pr(\text{state $i$}|z)$ that are comparably minor and almost negligible for moderately large covariate values. However, for Setting (II), with less persistence in the covariate process, there is a substantial bias across almost the entire covariate range --- which was to be expected, see Section \ref{sec:hypotheticalstadist}. In contrast, the AR resampling approach yields (visually) unbiased estimates for both (I) and (II).

\begin{figure}
    \centering
    \includegraphics[width=1\linewidth]{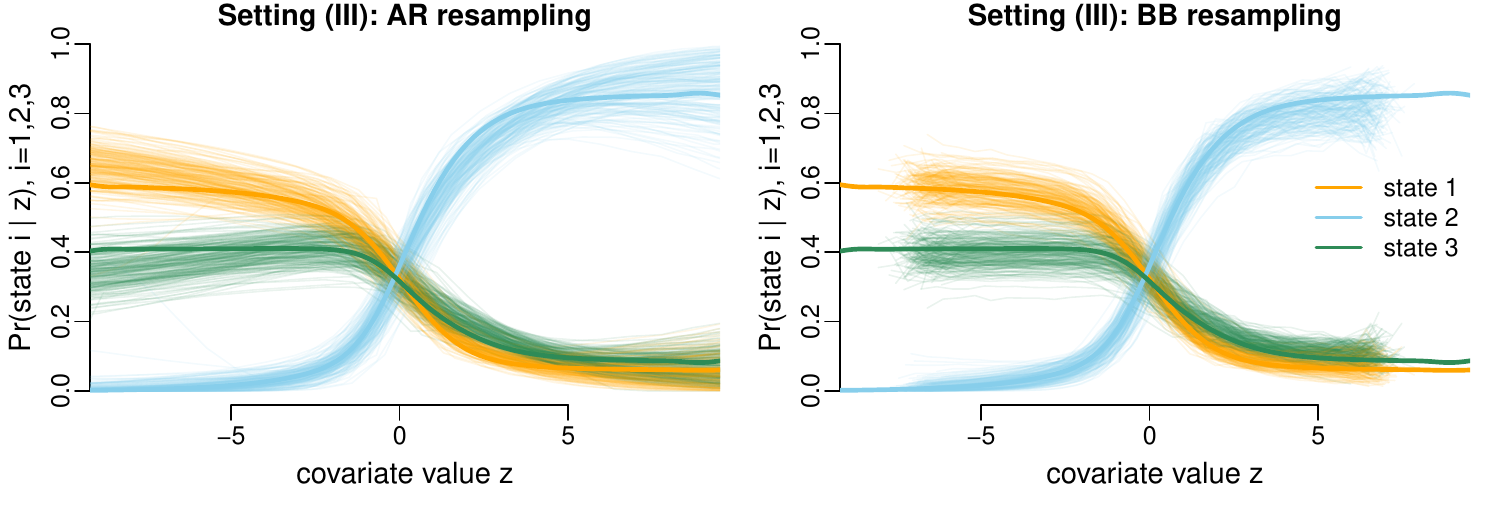}
    \caption{Comparison of AR and BB resampling for Setting (III) (periodic covariate).}
    \label{fig:simstudyARBB}
\end{figure}

In Setting (III), where the use of the AR resampling approach corresponds to a model misspecification of the covariate process, the approach fails to reproduce the seasonal pattern in the true process, which leads to a slightly biased estimate of $\Pr(\text{state $i$}|z)$ (cf.\ left panel in Figure \ref{fig:simstudyARBB}). The alternative, nonparametric approach based on the BB approach leads to (visually) unbiased, but wiggly estimates (cf.\ right panel in Figure \ref{fig:simstudyARBB}) --- these could of course still be smoothed, which we did not do here simply to illustrate the actual behaviour of the estimator. 

In summary, the simulation studies demonstrate that the potential bias resulting from the use of the hypothetical stationary distribution can be severe, with systematic under- and overestimation of the state occupancy probabilities of up to 30\% for certain covariate values in Setting (II).  The AR resampling approach produces reliable estimates when the true covariate process does have an autoregressive structure. When the latter is not the case, then the assumption-free BB approach can be preferable. The Dirichlet regression approach, the results of which are shown in the Online Supplementary Material (S.2), produces reliable estimates for at least moderately large sample sizes.



\section{Case study}
\label{sec:CaseStudy}

We consider GPS data collected from the \textit{Galápagos Tortoise Movement Ecology Programme}, publicly available through the Movebank Data Repository \citep{movebankgalapagos2019}
\footnote{\url{https://datarepository.movebank.org/handle/10255/move.834}}. 
As ectotherms, Galápagos tortoises (\textit{Chelonoidis niger}) regulate body temperature through environmental heat exchange rather than internal metabolic processes \citep{bastille2019migration}. Their activity therefore varies closely with ambient temperature: at cooler temperatures, they tend to move less and spend more time in sheltered areas, while warmer conditions encourage movement for foraging or exploration purposes. This physiological dependence makes temperature a natural candidate as a covariate affecting transitions between behavioural states, and we will specify the HMM accordingly below.

The dataset comprises hourly location records for 96 giant Galápagos tortoises. In the following, we focus on a single animal, ``Carolina", due to her relatively long monitoring period and low proportion of missing data. Carolina’s movement was recorded from 14 May 2009 to 29 October 2011, providing a detailed record of spatial and temporal variation in behaviour over more than two years. 
We preprocessed the data to obtain a regular hourly time series, removed outliers, and linearly imputed missing temperature observations --- the two longer periods with missing data (see Figure \ref{fig:CaseStudyTempTime}) were excluded from the analysis, with the overall HMM likelihood then taken to be the product of the likelihoods associated with the three separate extended periods with data collection. From the GPS coordinates, we derived the bivariate observation process of step lengths (in metres) and turning angles (in radians) \citep{michelot2016movehmm}. 

The model reported here, for illustration purposes, is a 3-state gamma/von Mises HMM as nowadays routinely used in movement ecology (see, e.g., \citealp{grecian2018understanding,beumer2020application}). The transition probabilities are modelled as functions of the ambient temperature, using a multinomial logistic link as in Eq.\ (\ref{eq:multilogitlinkTPM}), with linear predictors
$$
\eta_{ij}^{(t)} = \beta^{(ij)}_0 + \beta_1^{(ij)} \text{temp}_{t}, \quad i, j = 1, \dotsc, N, \quad i \neq j, 
$$
as to capture the tortoise’s thermally driven activity patterns. This model was fitted using numerical likelihood maximisation using the \texttt{LaMa} package in R \citep{lama}.

\begin{figure}[!htb]
  \centering
  \begin{subfigure}[t]{0.495\textwidth}
    \centering
    \includegraphics[width=\textwidth, keepaspectratio]{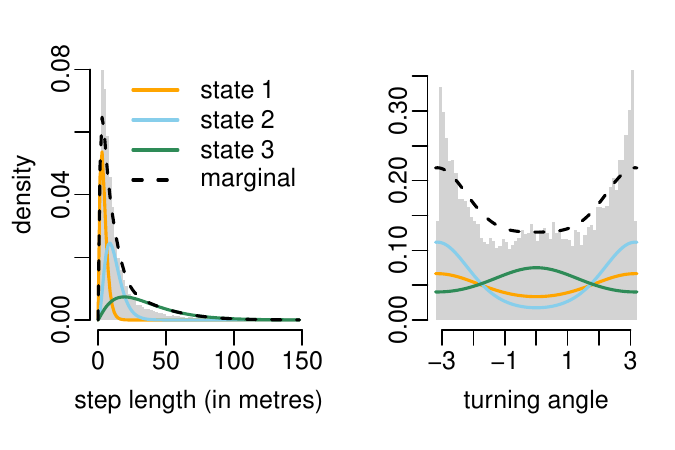}
    \caption{Estimated gamma and von Mises state-dependent distributions.}
    \label{fig:CaseStudyStateDependentDist}
  \end{subfigure}
  \hfill
  \begin{subfigure}[t]{0.495\textwidth}
    \centering
    \includegraphics[width=0.7\textwidth, keepaspectratio]{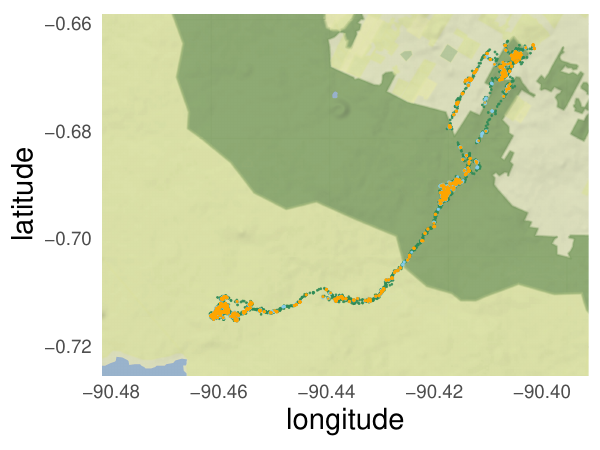}
    \caption{GPS track colour-coded according to the most likely state sequence.}
    \label{fig:CaseStudyTrack}
    \vspace{6pt}
  \end{subfigure}
  \vspace{4pt}
  \begin{subfigure}{\textwidth}
      \centering
      \includegraphics[width=0.9\textwidth,keepaspectratio]{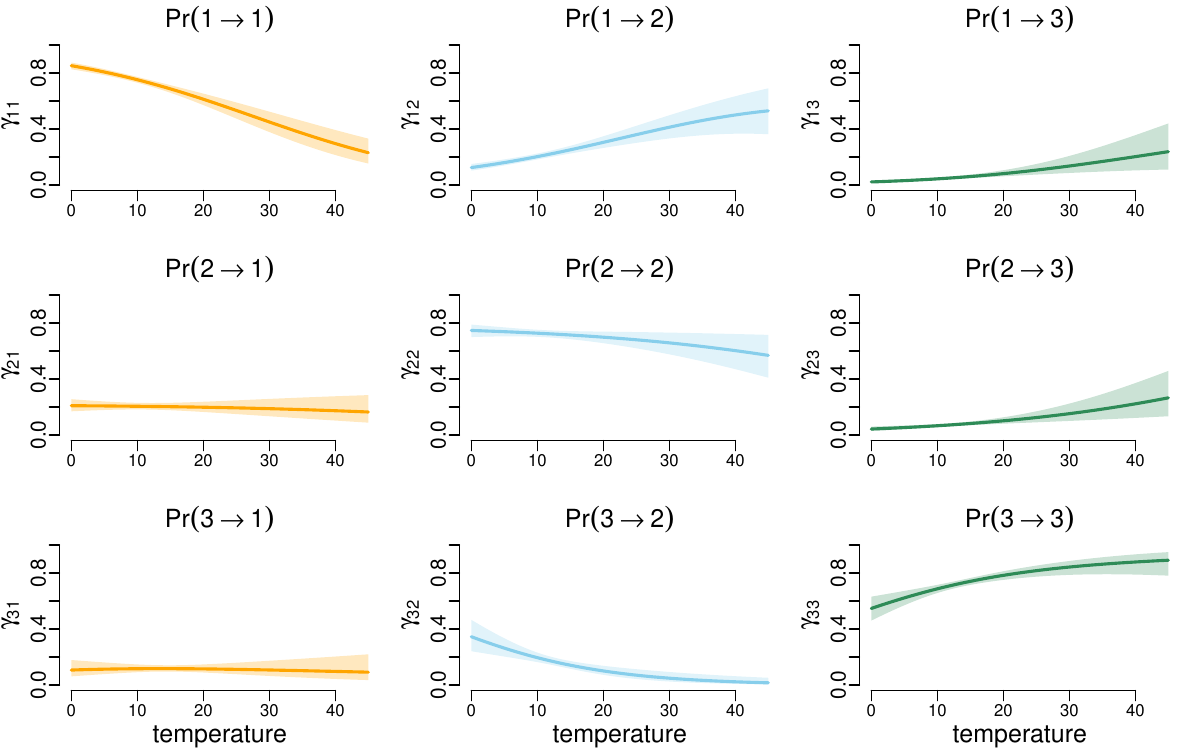}
      \caption{Estimated functional relationships between transition probabilities and temperature.}
      \label{fig:CaseStudyTranstionprob}
  \end{subfigure}
  \caption{Visualisations associated with the 3-state gamma/von Mises HMM fitted to the Galápagos tortoise's movement data.}
  \label{fig:CaseStudy4}
\end{figure}

Figure \ref{fig:CaseStudyStateDependentDist} displays the estimated state-dependent distributions. State 1 represents a low activity state with a mean step length of 4.3 metres and no directional persistence, State 2 corresponds to moderate movement activity (mean step length: 11.9 metres) and even higher tortuosity, and State 3 is associated with active movement behaviour (mean step length: 35.8 metres) and higher directional persistence. These states can loosely be interpreted as ``resting'', ``area-restricted search'' and ``travelling'', respectively, as also confirmed by the movement track colour-coded according to the most likely state sequence under the fitted HMM (Figure \ref{fig:CaseStudyTrack}).
The functional relationships between the temperature covariate and the transition probabilities (cf.\ Figure \ref{fig:CaseStudyTranstionprob}) reveal that as temperature increases, the probability of switching to the travelling mode (State 3) increases, while transitions to the resting mode (State 1) become less frequent.

\begin{figure}[!htb]
  \centering
  \begin{subfigure}[t]{0.49\textwidth}
    \centering
    \includegraphics[width=\textwidth, keepaspectratio]{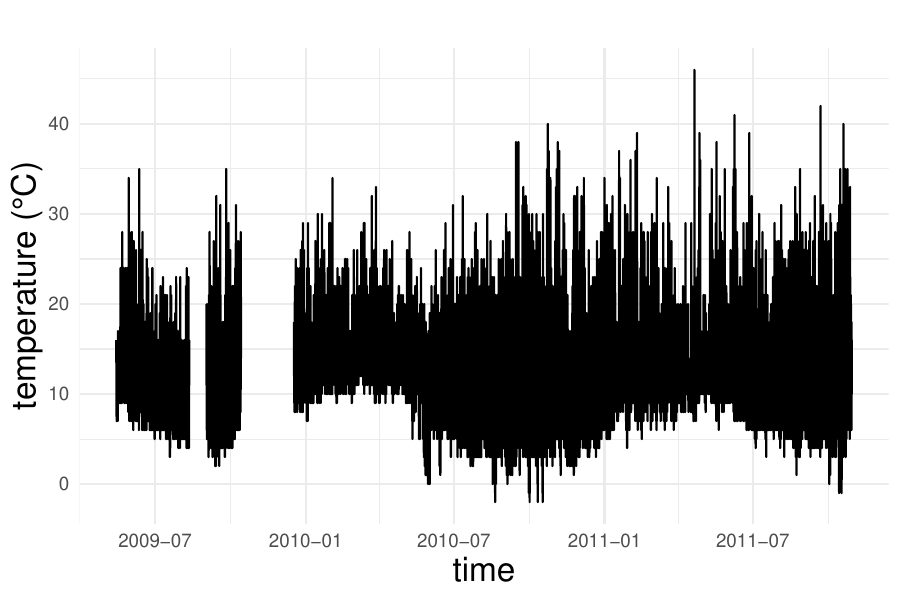}
    \caption{Time series of ambient temperature recorded during the tracking period.}
    \label{fig:CaseStudyTempTime}
  \end{subfigure}
  \hfill
  \begin{subfigure}[t]{0.49\textwidth}
    \centering
    \includegraphics[width=\textwidth, keepaspectratio]{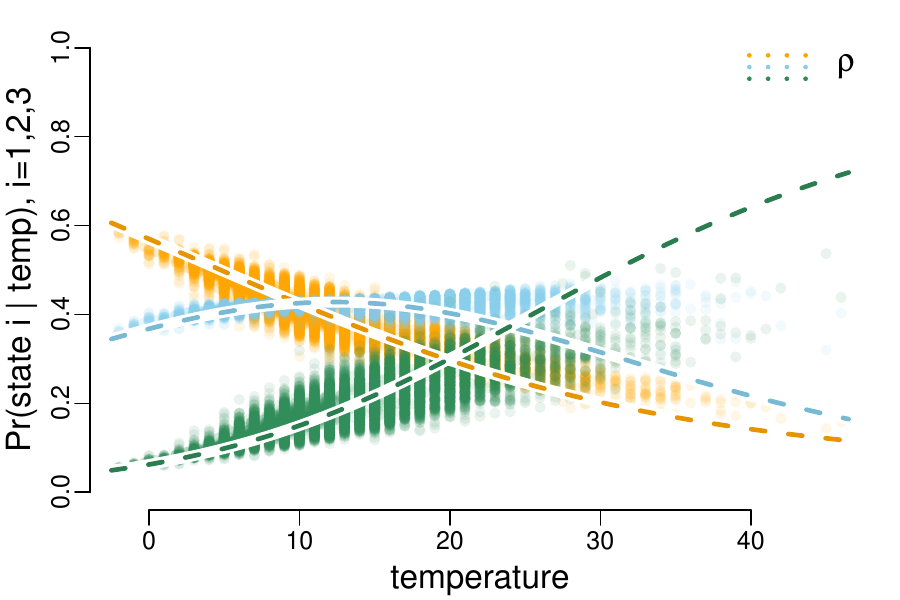}
    \caption{Hypothetical stationary distribution (dashed lines) and empirical state probabilities (points).}
    \label{fig:HypotheticalCS}
  \end{subfigure}
  \vspace{4pt}
  \begin{subfigure}[t]{0.49\textwidth}
    \centering
    \includegraphics[width=\textwidth, keepaspectratio]{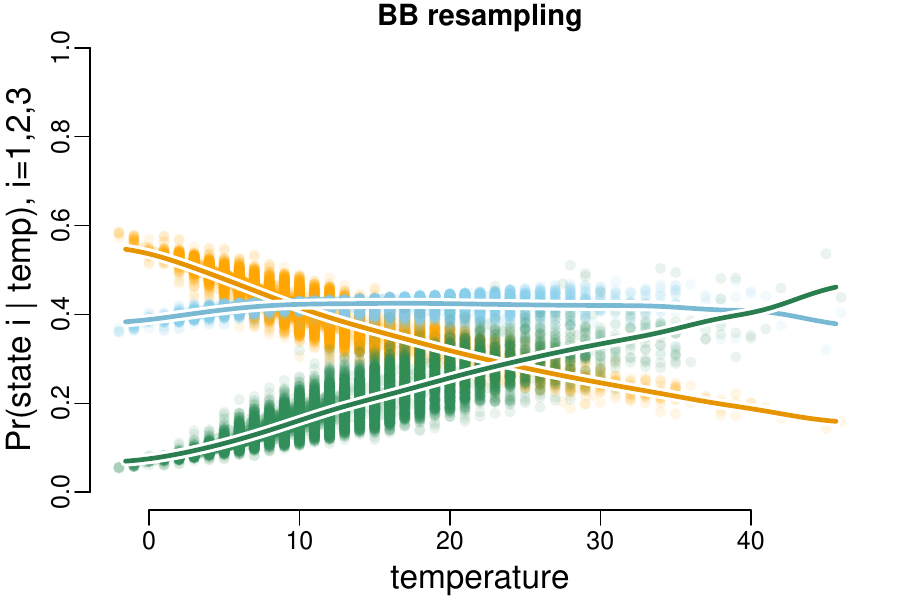}
    \caption{Covariate-dependent state occupancy inferred using BB (lines) and empirical state probabilities (points).}
    \label{fig:BBAppraochCS}
  \end{subfigure}
  \hfill
  \begin{subfigure}[t]{0.49\textwidth}
    \centering
    \includegraphics[width=\textwidth, keepaspectratio]{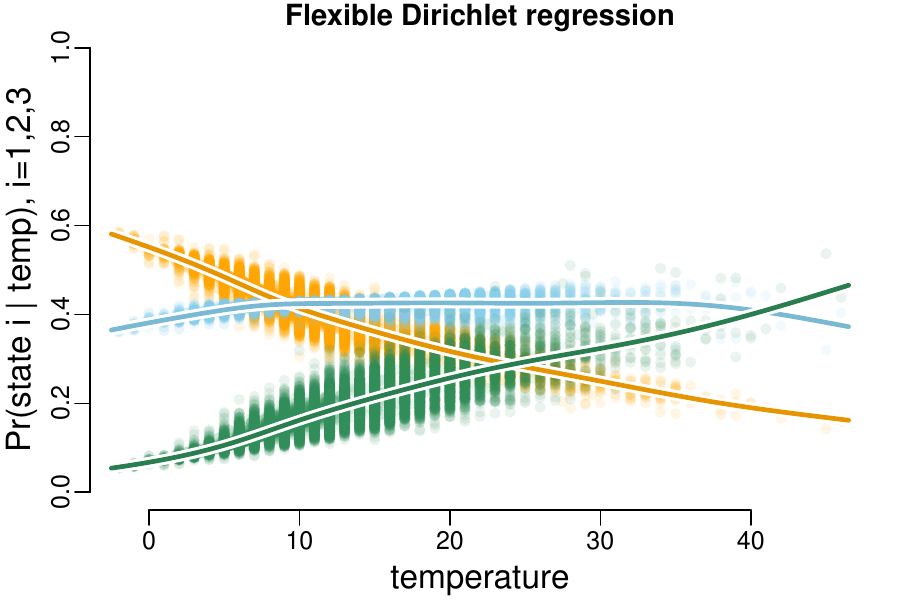}
    \caption{Covariate-dependent state occupancy inferred using Dirichlet regression (lines) and empirical state probabilities (points).}
    \label{fig:flexibledirichletRegS}
  \end{subfigure}
  \caption{Visualisations related to inference on covariate-driven state occupancy, under the fitted 3-state gamma/von Mises HMM.}
  \label{fig:flexibledirichletRegression}
\end{figure}

The time series of the temperature covariate is depicted in Figure \ref{fig:CaseStudyTempTime}, indicating both time-of-day and seasonal variation as well as positive skewness of the marginal distribution of the covariate.
Figure \ref{fig:HypotheticalCS} visualises the approximation of $\Pr(\text{state $i$}|z)$ obtained under the fitted model
when using the hypothetical stationary distribution, i.e.\ when ignoring the dynamics of the covariate process. In addition, this plot shows the model-implied state-occupancy probabilities $\bm{\delta}^{(t)}$ for the observed covariate values (see Section \ref{sec:directmodelling}), which indicate a bias of the hypothetical stationary distribution. The bias is strongest for large temperature values, likely due to the skewness of the temperature distribution, with very large values obtained only rarely and for brief periods.

As alternative to the approximation using the hypothetical stationary distribution, we implement both the BB approach (Section \ref{sec:resampling-approaches}) and Dirichlet regression (Section \ref{sec:directmodelling}) --- (basic) AR processes are not as suitable here given the skewness and the periodic dynamics in the covariate process. For the BB approach, we first model the seasonal variation using the regression model
$$
\text{temp}_{t} = \beta_0 + \beta_1 \sin \Bigl( \frac{2 \pi \text{doy}_t}{365} \Bigr) + \beta_2\cos \Bigl( \frac{2 \pi \text{doy}_t}{365} \Bigr)+\epsilon_t,
$$
where $\text{doy}_t$ denotes the day of the year at observation index $t$. We then use blocks of residuals $\epsilon_t$ of length 24 in the BB approach, which after resampling are again added to the estimated trigonometric function above to produce realistic simulated temperature trajectories. For the approach based on Dirichlet regression, we use thin-plate regression splines as the smoothing basis.

The results obtained under the BB and the Dirichlet regression approach can be found in Figures \ref{fig:BBAppraochCS} and \ref{fig:flexibledirichletRegS}, respectively; a version also including confidence intervals is provided in the Online Supplementary Material (S.3). Both approaches yield estimates of $\Pr(\text{state $i$}|z)$ that closely follow the corresponding empirical quantities, with no indication of any bias. In either case, the main ecological finding here is that higher temperatures substantially increase the proportion of time the tortoise spends in the travelling mode, whereas resting behaviour is most frequently observed at cold temperatures.

\section{Discussion}

In analyses of behavioural time series data using covariate-driven HMMs, resulting biological inference is often based on the implied covariate-dependent state occupancy distribution (see, e.g., \citealp{nagel2021movement}, \citealp{fahlbusch2022blue}, \citealp{whittington2022towns},  \citealp{ferreiro2024drivers}). The latter is not analytically available and has to date commonly been approximated using the hypothetical stationary distribution as suggested by \citet{patterson2009classifying}.
As documented here, this approach can lead to substantial biases, especially if the covariate fluctuates considerably over time. It should be noted that the authors of \citet{patterson2009classifying} presumably never intended their approach to become the default option for approximating the covariate-dependent state occupancy distribution: in particular, in their case, the covariate considered, namely the sea surface temperature, was in fact highly persistent, such that the bias resulting from the use of the hypothetical stationary distribution will have been negligible. 

For settings in which the covariate process does exhibit notable volatility, we presented three different approaches to arrive at improved and ideally unbiased estimates of the covariate-dependent state occupancy distribution. We recommend the use of resampling-based approaches in scenarios with relatively short time series, say in the order of up to a few thousand observations, such that there will typically be insufficient information on the covariates' distribution and dynamics to directly estimate $\Pr(\text{state $i$}|z)$ from its empirical counterpart. The approach using autoregressive process modelling then constitutes a relatively robust default option that requires very little input by the analyst, but should be replaced by the block bootstrap approach when there is reason to believe that the AR model cannot adequately capture the actual dynamics of the covariate process. In settings with relatively long time series (several thousand or more observations), the empirical versions of $\Pr(\text{state $i$}|z)$ will typically provide sufficient information for fitting a smooth curve, such that in those cases the Dirichlet regression approach will usually be a very good option that again requires little input by the analyst. All three approaches naturally extend to the case of multiple covariates.

\printbibliography

@incollection{visser2017markov,
  title={Markov process models for discrimination learning},
  author={Visser, Ingmar and Schmittmann, Verena D and Raijmakers, Maartje EJ},
  booktitle={Longitudinal Models in the Behavioral and Related Sciences},
  pages={337--366},
  year={2017},
  publisher={Routledge}
}

@article{mews2025build,
  title={How to build your latent {M}arkov model: The role of time and space},
  author={Mews, Sina and Koslik, Jan-Ole and Langrock, Roland},
  journal={Statistical Modelling},
  volume={1471082X251355681},
  year={2025},
  publisher={SAGE Publications Sage India: New Delhi, India}
}

@article{nagel2021movement,
  title={Movement patterns and activity levels are shaped by the neonatal environment in {A}ntarctic fur seal pups},
  author={Nagel, Rebecca and Mews, Sina and Adam, Timo and Stainfield, Claire and Fox-Clarke, Cameron and Toscani, Camille and Langrock, Roland and Forcada, Jaume and Hoffman, Joseph I},
  journal={Scientific Reports},
  volume={11},
  number={1},
  pages={14323},
  year={2021},
  publisher={Nature Publishing Group UK London}
}

@book{zucchini,
  title={Hidden {M}arkov {M}odels for {T}ime {S}eries: {A}n {I}ntroduction {U}sing {R}},
  author={Zucchini, Walter and MacDonald, Iain L and Langrock, Roland},
  year={2016},
  publisher={Boca Raton: Chapman \& Hall/CRC}
}

@article{ferreiro2024drivers,
  title={Drivers of wolf activity in a human-dominated landscape and its individual variability toward anthropogenic disturbance},
  author={Ferreiro-Arias, Iago and Garc{\'\i}a, Emilio Jos{\'e} and Palacios, Vicente and Sazatornil, V{\'\i}ctor and Rodr{\'\i}guez, Alejandro and L{\'o}pez-Bao, Jos{\'e} Vicente and Llaneza, Luis},
  journal={Ecology and Evolution},
  volume={14},
  number={10},
  pages={e70397},
  year={2024},
  publisher={Wiley Online Library}
}

@article{fahlbusch2022blue,
  title={Blue whales increase feeding rates at fine-scale ocean features},
  author={Fahlbusch, James A and Czapanskiy, Max F and Calambokidis, John and Cade, David E and Abrahms, Briana and Hazen, Elliott L and Goldbogen, Jeremy A},
  journal={Proceedings of the Royal Society B},
  volume={289},
  number={1981},
  pages={20221180},
  year={2022},
  publisher={The Royal Society}
}

@article{whittington2022towns,
  title={Towns and trails drive carnivore movement behaviour, resource selection, and connectivity},
  author={Whittington, Jesse and Hebblewhite, Mark and Baron, Robin W and Ford, Adam T and Paczkowski, John},
  journal={Movement Ecology},
  volume={10},
  number={1},
  pages={17},
  year={2022},
  publisher={Springer}
}

@article{feldmann2023flexible,
  title={Flexible modelling of diel and other periodic variation in hidden {M}arkov models},
  author={Feldmann, Carlina C and Mews, Sina and Coculla, Angelica and Stanewsky, Ralf and Langrock, Roland},
  journal={Journal of Statistical Theory and Practice},
  volume={17},
  number={3},
  pages={45},
  year={2023},
  publisher={Springer}
}

@article{towner2016sex,
  title={Sex-specific and individual preferences for hunting strategies in white sharks},
  author={Towner, Alison V and Leos-Barajas, Vianey and Langrock, Roland and Schick, Robert S and Smale, Malcolm J and Kaschke, Tami and Jewell, Oliver JD and Papastamatiou, Yannis P.},
  journal={Functional Ecology},
  volume={30},
  number={8},
  pages={1397--1407},
  year={2016},
  publisher={Wiley Online Library}
}

@article{klappstein2023flexible,
  title={Flexible hidden {M}arkov models for behaviour-dependent habitat selection},
  author={Klappstein, Natasha Jean and Thomas, Len and Michelot, Th{\'e}o},
  journal={Movement Ecology},
  volume={11},
  number={1},
  pages={30},
  year={2023},
  publisher={Springer}
}

@article{creel2020hidden,
  title={Hidden {M}arkov models reveal a clear human footprint on the movements of highly mobile {A}frican wild dogs},
  author={Creel, Scott and Merkle, Johnathan and Mweetwa, Thandiwe and Becker, Matthew S and Mwape, Henry and Simpamba, Twakundine and Simukonda, Chuma},
  journal={Scientific Reports},
  volume={10},
  number={1},
  pages={17908},
  year={2020},
  publisher={Nature Publishing Group UK London}
}

@article{grecian2018understanding,
  title={Understanding the ontogeny of foraging behaviour: insights from combining marine predator bio-logging with satellite-derived oceanography in hidden {M}arkov models},
  author={Grecian, James W and Lane, Jude V and Michelot, Th{\'e}o and Wade, Helen M and Hamer, Keith C},
  journal={Journal of the Royal Society Interface},
  volume={15},
  number={143},
  pages={20180084},
  year={2018},
  publisher={The Royal Society}
}

@article{beumer2020application,
  title={An application of upscaled optimal foraging theory using hidden {M}arkov modelling: year-round behavioural variation in a large arctic herbivore},
  author={Beumer, Larissa T and Pohle, Jennifer and Schmidt, Niels M and Chimienti, Marianna and Desforges, Jean-Pierre and Hansen, Lars H and Langrock, Roland and Pedersen, Stine H{\o}jlund and Stelvig, Mikkel and van Beest, Floris M},
  journal={Movement Ecology},
  volume={8},
  number={1},
  pages={25},
  year={2020},
  publisher={Springer}
}

@article{byrnes2021evaluating,
  title={Evaluating the constraints governing activity patterns of a coastal marine top predator},
  author={Byrnes, Evan E and Daly, Ryan and Leos-Barajas, Vianey and Langrock, Roland and Gleiss, Adrian C},
  journal={Marine Biology},
  volume={168},
  number={1},
  pages={11},
  year={2021},
  publisher={Springer}
}

@article{farhadinia2020understanding,
  title={Understanding decision making in a food-caching predator using hidden {M}arkov models},
  author={Farhadinia, Mohammad S and Michelot, Th{\'e}o and Johnson, Paul J and Hunter, Luke TB and Macdonald, David W},
  journal={Movement Ecology},
  volume={8},
  number={1},
  pages={9},
  year={2020},
  publisher={Springer}
}

@article{koslik2023inference,
  title={Inference on the state process of periodically inhomogeneous hidden {M}arkov models for animal behavior},
  author={Koslik, Jan-Ole and Feldmann, Carlina C and Mews, Sina and Michels, Rouven and Langrock, Roland},
  journal={The {A}nnals of {A}pplied {S}tatistics},
  volume={19},
  number={4},
  pages={2724--2737},
  year={2025},
  publisher={{I}nstitute of {M}athematical {S}tatistics}
}

@article{jackson2003multistate,
  title={Multistate {M}arkov models for disease progression with classification error},
  author={Jackson, Christopher H and Sharples, Linda D and Thompson, Simon G and Duffy, Stephen W and Couto, Elisabeth},
  journal={Journal of the Royal Statistical Society Series D: The Statistician},
  volume={52},
  number={2},
  pages={193--209},
  year={2003},
  publisher={Oxford University Press}
}

@Manual{lama,
 title = {{LaMa}: {F}ast Numerical Maximum Likelihood Estimation for Latent {M}arkov Models},
 author = {Jan-Ole Koslik},
 year = {2025},
 note = {R package version 2.0.6},
 url = {https://CRAN.R-project.org/package=LaMa}
}

@article{mcclintock2020uncovering,
  title={Uncovering ecological state dynamics with hidden {M}arkov models},
  author={McClintock, Brett T and Langrock, Roland and Gimenez, Olivier and Cam, Emmanuelle and Borchers, David L and Glennie, Richard and Patterson, Toby A},
  journal={Ecology Letters},
  volume={23},
  number={12},
  pages={1878--1903},
  year={2020},
  publisher={Wiley Online Library}
}

@article{van2019classifying,
  title={Classifying grey seal behaviour in relation to environmental variability and commercial fishing activity-a multivariate hidden {M}arkov model},
  author={{van Beest}, Floris M and Mews, Sina and Elkenkamp, Svenja and Schuhmann, Patrick and Tsolak, Dorian and Wobbe, Till and Bartolino, Valerio and Bastardie, Francois and Dietz, Rune and von Dorrien, Christian and others},
  journal={Scientific Reports},
  volume={9},
  number={1},
  pages={5642},
  year={2019},
  publisher={Nature Publishing Group UK London}
}

@article{patterson2017statistical,
  title={Statistical modelling of individual animal movement: an overview of key methods and a discussion of practical challenges},
  author={Patterson, Toby A and Parton, Alison and Langrock, Roland and Blackwell, Paul G and Thomas, Len and King, Ruth},
  journal={AStA Advances in Statistical Analysis},
  volume={101},
  number={4},
  pages={399--438},
  year={2017},
  publisher={Springer}
}

@article{patterson2009classifying,
  title={Classifying movement behaviour in relation to environmental conditions using hidden {M}arkov models},
  author={Patterson, Toby A and Basson, Marinelle and Bravington, Mark V and Gunn, John S},
  journal={Journal of Animal Ecology},
  volume={78},
  number={6},
  pages={1113--1123},
  year={2009},
  publisher={Wiley Online Library}
}

@article{douma2019analysing,
  title={Analysing continuous proportions in ecology and evolution: A practical introduction to beta and {D}irichlet regression},
  author={Douma, Jacob C and Weedon, James T},
  journal={Methods in Ecology and Evolution},
  volume={10},
  number={9},
  pages={1412--1430},
  year={2019},
  publisher={Wiley Online Library}
}

@misc{movebankgalapagos2019,
	author = {Bastille-Rousseau, Guillaume and Yackulic, CB and Gibbs, J and Frair, JL and Cabrera, F and Blake, S},
	doi = {doi:10.5441/001/1.6gr485fk},
	publisher = {Movebank data repository},
	title = {Data from: Migration triggers in a large herbivore: Gal{\'a}pagos giant tortoises navigating resource gradients on volcanoes},
	url = {http://dx.doi.org/10.5441/001/1.6gr485fk},
	year = {2019}}

@article{mcclintock2018momentuhmm,
  title={{momentuHMM}: {R} package for generalized hidden {M}arkov models of animal movement},
  author={McClintock, Brett T and Michelot, Th{\'e}o},
  journal={Methods in Ecology and Evolution},
  volume={9},
  number={6},
  pages={1518--1530},
  year={2018},
  publisher={Wiley Online Library}
}

@article{michelot2016movehmm,
  title={{moveHMM}: an {R} package for the statistical modelling of animal movement data using hidden {M}arkov models},
  author={Michelot, Th{\'e}o and Langrock, Roland and Patterson, Toby A},
  journal={Methods in Ecology and Evolution},
  volume={7},
  number={11},
  pages={1308--1315},
  year={2016},
  publisher={Wiley Online Library}
}

@article{dahl2022time,
  title={Time series (re) sampling using generative adversarial networks},
  author={Dahl, Christian M and S{\o}rensen, Emil N},
  journal={Neural Networks},
  volume={156},
  pages={95--107},
  year={2022},
  publisher={Elsevier}
}

@article{goodfellow2014generative,
  title={Generative adversarial nets},
  author={Goodfellow, Ian J and Pouget-Abadie, Jean and Mirza, Mehdi and Xu, Bing and Warde-Farley, David and Ozair, Sherjil and Courville, Aaron and Bengio, Yoshua},
  journal={Advances in Neural Information Processing Systems},
  volume={27},
  year={2014}
}

@article{kunsch1989jackknife,
  title={The jackknife and the bootstrap for general stationary observations},
  author={K{\"u}nsch, Hans R},
  journal={The Annals of Statistics},
  pages={1217--1241},
  year={1989},
  publisher={JSTOR}
}

@article{ives2010analysis,
  title={Analysis of ecological time series with ARMA (p, q) models},
  author={Ives, Anthony R and Abbott, Karen C and Ziebarth, Nicolas L},
  journal={Ecology},
  volume={91},
  number={3},
  pages={858--871},
  year={2010},
  publisher={Wiley Online Library}
}

@article{koslik2024efficientsmoothnessselectionnonparametric,
      title={Efficient smoothness selection for nonparametric {M}arkov-switching models via quasi restricted maximum likelihood}, 
      author={Jan-Ole Koslik},
      year={2024},
      eprint={2411.11498},
      archivePrefix={arXiv}
}

@article{bastille2019migration,
  title={Migration triggers in a large herbivore: Gal{\'a}pagos giant tortoises navigating resource gradients on volcanoes},
  author={Bastille-Rousseau, Guillaume and Yackulic, Charles B and Gibbs, James P and Frair, Jacqueline L and Cabrera, Freddy and Blake, Stephen},
  journal={Ecology},
  volume={100},
  number={6},
  pages={e02658},
  year={2019},
  publisher={Wiley Online Library}
}

\end{document}